\documentclass[prb,amsmath,amssymb,twocolumn,showpacs,floatfix]{revtex4}
\usepackage{graphicx}
\usepackage{dcolumn}
\usepackage{bm}

\begin{document}

\title{Detection of precursor charge-density-wave order in 2$H$-NbSe$_2$ by $\mu$SR}

\author{C.~Gomez,$^1$ D.~Braam,$^1$ S.~Tezok,$^1$ and J.E.~Sonier,$^{1,2}$}

\affiliation{$^1$Department of Physics, Simon Fraser University, Burnaby, British Columbia V5A 1S6, Canada \\
$^2$Canadian Institute for Advanced Research, Toronto, Ontario M5G 1Z8, Canada}

\date{\today}

\begin{abstract}
We demonstrate the sensitivity of transverse-field muon spin rotation (TF-$\mu$SR) to static charge-density-wave (CDW) order
in the bulk of 2$H$-NbSe$_2$. In the presence of CDW order the quadrupolar interaction of the $^{93}$Nb nuclei with the local electric-field
gradient is modified, and this in turn affects the magnetic dipolar coupling of the positive muon to these nuclei.
For a weak magnetic field applied parallel to the $c$-axis, we observe a small enhancement of the muon depolarization
rate at temperatures below the established CDW phase transition. Aligning the applied field perpendicular to the $c$-axis,
we observe a sensivity to static CDW order in regions of the sample extending up to nearly 3 times the CDW transition 
temperature ($T_0$). The results suggest that the muon is mobile over the temperature range explored above the superconducting 
transition temperature ($T_c$), and becomes trapped in the vicinity of defects.     
\end{abstract}


\maketitle

\section{Introduction}

There is currently much interest in the interplay of CDW and superconducting states, driven by the discovery of
short-range CDW correlations in the pseudogap phase of cuprates.\cite{Chang:12,Ghiringhelli:12,Comin:14,Neto:14,Blanco:14,Huecker:14,Neto:15} 
Upon lowering the temperature the CDW correlation volume expands, but a reversal of this trend below $T_c$ indicates a clear
competition of the CDW order with superconductivity.
Competing CDW order and superconductivity has also been demonstrated in the transition-metal dichalcogenide 2$H$-NbSe$_2$,\cite{Borisenko:09} 
which like the cuprates is quasi two-dimensional. With decreasing temperature, 2$H$-NbSe$_2$ undergoes an incommensurate 
CDW transition at $T_0 \! \sim \! 33$~K and becomes superconducting below $T_c \! \sim \! 7$~K.\cite{Wilson:75}
In contrast with cuprates, superconductivity in 2$H$-NbSe$_2$ does not coexist with magnetism or lie in close proximity to
a magnetic phase, and is mediated by a conventional electron-phonon interaction.

The occurrence of a long-range CDW state in 2$H$-NbSe$_2$ for an extensive range of temperature above $T_c$ differs from the 
situation in cuprates, where only short-range CDW correlations are observed above $T_c$. 
On the other hand, $^{93}$Nb nuclear magnetic resonance (NMR) measurements show pre-transitional broadening of 
quadrupolar transitions below $T \! \sim \! 75$~K, which has been attributed to precursor CDW fluctuations.\cite{Stiles:76,Berthier:78,Ghoshray:09} 
Berthier {\it et al.} proposed that the CDW fluctuations inferred from the NMR measurements are induced near impurities,\cite{Berthier:78}
which bears some resemblance to a more recent proposal by Tacon {\it et al.} that CDW correlations in cuprates are
pinned in nanometer-size regions by lattice defects.\cite{Tacon:14} More recently, nanodomains of static CDW order have been observed 
near defects in 2$H$-NbSe$_2$ at temperatures $T_0 \! < \! T \! < \! 100$~K by scanning tunneling microscopy (STM).\cite{Arguello:14}
Intercalating 2$H$-NbSe$_2$ with Co or Mn ions reduces $T_0$, but short-range CDW order is found to persist for an extensive range of
temperature above $T_0$.\cite{Chatterjee:15} 
 
The short-lived (with a mean lifetime of approximately 2.2~$\mu$s), spin $S \! = \! 1/2$, positive muon ($\mu^+$) 
is most often used in condensed matter physics as a pure local magnetic probe, since it does not possess an electric quadrupole moment.
However, magnetic dipolar coupling to quadrupolar nuclei provides an indirect sensitivity of the $\mu^+$ to CDW correlations in the bulk.
In 2$H$-NbSe$_2$ the spin of the implanted $\mu^+$ dipolar couples to the $^{93}$Nb and $^{77}$Se nuclear magnetic moments.
The $^{93}$Nb nuclei have spin $I^{\rm Nb} \! = \! 9/2$ and possess an electric quadrupole moment ($^{77}$Se, which has
nuclear spin $I^{\rm Se} \! = \! 1/2$, does not), which 
interacts with the local electric-field gradient (EFG). In the presence of an applied static magnetic field {\bf B}$_0$, both the $\mu^+$
and $^{77}$Se spins precess around {\bf B}$_0$, the direction of which serves as a suitable quantization axis. 
On the other hand, the quantization axis for the $^{93}$Nb nuclear spins is oriented along the vector 
sum of {\bf B}$_0$ and the direction of the maximal local EFG. 

In a TF-$\mu$SR experiment, where the applied magnetic field {\bf B}$_0$ (chosen to be along the $z$ direction)
is perpendicular to the initial muon spin polarization {\bf P}(0) (chosen here to be along the $x$ direction), the time evolution of the muon 
spin polarization for 2$H$-NbSe$_2$ is well described by
\begin{eqnarray}
P_x(t) & = & \exp(-\sigma^2 t^2) \cos(\gamma_{\mu} B_{\mu} t + \phi) \nonumber \\
& = & \exp[-(\sigma_{\rm Nb}^2 + \sigma_{\rm Se}^2)t^2]\cos(\gamma_{\mu} B_{\mu} t + \phi) \, ,
\label{eq:Pol}
\end{eqnarray}
where $\gamma_{\mu}$ is the muon gyromagnetic ratio, $B_{\mu}$ is the mean magnetic field
sensed by the muon ensemble, $\phi$ is a phase constant, and $\sigma_{\rm Nb}$ ($\sigma_{\rm Se}$)
is the Gaussian depolarization rate associated with the dipolar coupling of the $\mu^+$ to the $^{93}$Nb ($^{77}$Se) nuclei.
The muon depolarization rates are related to the width of the distribution of nuclear dipole fields in the $z$ direction 
at the $\mu^+$ site via the relation $\sigma^2 \! = \! \gamma_{\mu} \langle B_{\mu, z}^2 \rangle$, and are given by
\begin{equation}
\sigma_{\rm Se}^2 = \frac{1}{3} \gamma_{\mu}^2 \gamma_{\rm Se}^2  \hbar^2 I^{\rm Se}(I^{\rm Se} + 1) \sum\limits_{i}
\frac{(1 - 3 \cos^2 \theta_i)}{r_i^6}
\label{eq:Se}
\end{equation}
and,\cite{Hartmann:77}
\begin{eqnarray}
\sigma_{\rm Nb}^2   = \gamma_{\mu}^2 \gamma_{\rm Nb}^2 \hbar^2 \sum\limits_{j} & \left\langle \left( \langle I_z^{\rm Nb} \rangle_j
\frac{(1 - 3 \cos^2 \theta_j)}{r_j^3} \right. \right. \nonumber \\ 
& \left. \left. - \langle I_x^{\rm Nb} \rangle_j \frac{3 \sin \theta_j \cos \theta_j}{r_j^3} \right)^2 \right\rangle_{\rm av} \, .
\label{eq:Nb}
\end{eqnarray}
Here $\gamma_{\rm Se}$ ($\gamma_{\rm Nb}$) is the gyromagnetic ratio of the $^{77}$Se ($^{93}$Nb) nuclei,
$r_i$ ($r_j$) the distance between the $\mu^+$ and the $i^{\rm th}$ $^{77}$Se ($j^{\rm th}$ $^{93}$Nb) nucleus, 
$\theta_i$ ($\theta_j$) is the angle between {\bf B}$_0$ and the straight line connecting the $\mu^+$ spin
and the $i^{\rm th}$ $^{77}$Se ($j^{\rm th}$ $^{93}$Nb) nuclear spin, $I^{\rm Se} \! = \! 1/2$ is the nuclear spin of $^{77}$Se,
$\langle I_z^{\rm Nb} \rangle_j$ and $\langle I_x^{\rm Nb} \rangle_j$ are the expectation values of the $^{93}$Nb
nuclear spin parallel and perpendicular to the $z$ direction, and $\langle \; \rangle_{\rm av}$ is the average
over all eigenstates of the $j^{\rm th}$ $^{93}$Nb nuclear spin. Note that in addition to the static spin
component $\langle I_z^{\rm Nb} \rangle_j$ parallel to {\bf B}$_0$, the quadrupole interaction of the $^{93}$Nb 
with the local EFG results in a static spin component $\langle I_x^{\rm Nb} \rangle_j$ perpendicular to {\bf B}$_0$.
The second term in Eq.~(\ref{eq:Nb}) then represents the contribution of this perpendicular static spin component 
to the dipole field in the $z$ direction at the $\mu^+$ site. Since the natural abundance of $^{77}$Se is only $\sim \! 7.6$~\%, 
the dipolar interaction with the $^{93}$Nb nuclei dominates the depolarization of the TF-$\mu$SR spectrum.     

The ratio $\langle I_x^{\rm Nb} \rangle_j/\langle I_z^{\rm Nb} \rangle_j$ depends on the ratio of the
quadrupolar and Zeeman interaction energies.
For strong magnetic field the Zeeman interaction of the $^{93}$Nb spin with {\bf B}$_0$ dominates and the quantization axis
of the $^{93}$Nb nuclear spins is along the $z$ direction. In this case $\langle(\langle I_x^{\rm Nb} \rangle_j)^2 \rangle_{\rm av} \! = \! 0$ 
and $\langle (\langle I_z^{\rm Nb} \rangle_j)^2 \rangle_{\rm av} \! = \! I(I+1)/3$, which reduces Eq.~(\ref{eq:Nb}) to an equation analogous to
Eq.~(\ref{eq:Se}). Of interest here is the situation at sufficiently low fields, where the magnitude of the quadrupolar interaction 
of the $^{93}$Nb nuclei with the local EFG exceeds or becomes comparable to the Zeeman interaction.
The quadrupolar coupling constant for $^{93}$Nb in $2H$-NbSe$_2$ above $T_0$ is $e^2qQ/h \! \sim \! 62$~MHz,\cite{Berthier:78}
and hence the $^{93}$Nb quadrupolar frequency is $\nu_{\rm Q} \! = \! 3e^2qQ/h2I(2I - 1) \! \sim \! 2.58$~MHz.
Consequently, the Zeeman and quadrupolar interaction energies are equivalent when the Zeeman frequency is $\nu_{\rm Z} \! = \! 2.58$~MHz,
which corresponds to a magnetic field of magnitude $B_0 \! = \! 2 \pi \nu_{\rm Z} / \gamma_{\rm Nb} \! = \! 2.47$~kG.
The EFG at the $^{93}$Nb nuclear site in $2H$-NbSe$_2$ is due to the non-cubic ionic crystal lattice, but
is modified by the presence of the $\mu^+$ and its screening charge, and by the development of CDW correlations. 

Since the $\mu$SR technique does not require the application of a magnetic field, greater sensitivity of the $\mu^+$
to CDW order may be achieved by removal of the Zeeman interaction. However, the zero-field (ZF) $\mu$SR spectra for 2$H$-NbSe$_2$ are
rather complex. Higemoto {\it et al.} showed that the ZF-$\mu$SR signals fit well to a two-component dynamical Kubo-Toyabe 
function, but the physical interpretation of the various fit parameters are ambiguous.\cite{Higemoto:99}
Here we show via TF-$\mu$SR measurements of 2$H$-NbSe$_2$ that the complexity of the ZF-$\mu$SR spectra are the result 
of muon diffusion. Unlike the situation for zero field, the TF-$\mu$SR signals are well described by a simple 
single-component depolarization function, which greatly simplifies the interpretation of the data.  

\begin{figure}
\centering
\includegraphics[width=8.0cm]{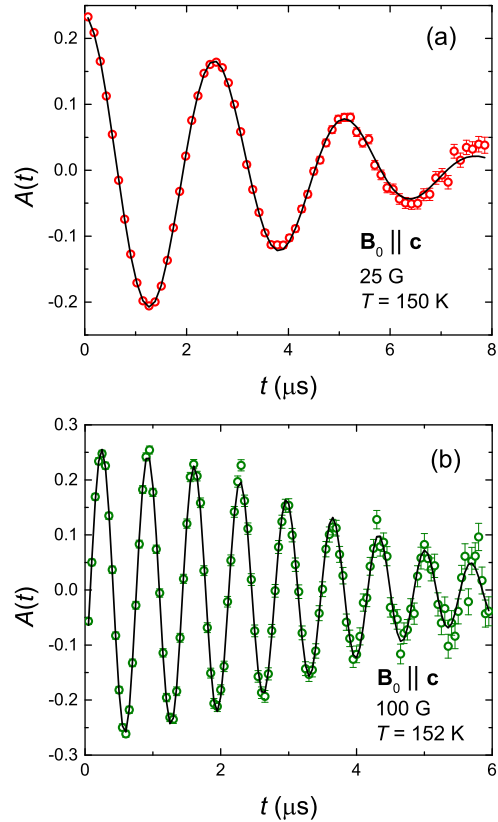}
\caption{(Color online) Representative TF-$\mu$SR spectra recorded at a temperature near 150~K and for a magnetic field of 
(a) 25~G, and (b) 100~G applied parallel to the $c$ axis of 2$H$-NbSe$_2$. The solid curves through the data points are
fits to Eq.~(\ref{eq:TotalPol}).}
\label{fig1}
\end{figure}

\section{Experimental Details}

Our TF-$\mu$SR experiments were carried out on a previously studied single crystal of 2$H$-NbSe$_2$,\cite{Callaghan:05} 
which has a superconducting transition temperature of $T_c \! = \! 7.0$~K and an upper critical magnetic field of $B_{c2} \! = \! 45$~kG.
The LAMPF and HELIOS spectrometers were used on the M20 surface $\mu^+$ beam line at TRIUMF in Vancouver, Canada.
The sample was mounted with the $c$ axis parallel to the $\mu^+$-beam linear momentum. The majority of the measurements were
performed with the applied magnetic field {\bf B}$_0$ parallel to the $c$ axis, since the Lorentz force exerted by a perpendicular 
field causes deflection of the muon beam. Consequently, measurements with {\bf B}$_0$ perpendicular to the $c$ axis were
limited to the applied field strength of 25~G. In all cases the initial muon spin polarization {\bf P}(0)
was oriented perpendicular to {\bf B}$_0$.

The TF-$\mu$SR spectra were fit to
\begin{equation}
A(t) = a_s P(t) + a_b \exp(-\sigma_b^2 t^2/2) \cos(\gamma_{\mu} B_0 t + \varphi) \, ,
\label{eq:TotalPol}
\end{equation}
where $a_s$ and $a_b$ are the initial amplitudes of the sample and background signals, $P(t)$ is as defined
in Eq.~(\ref{eq:Pol}), and $\sigma_b$ and $\varphi$ are the muon depolarization rate and phase constant
of the background component associated with muons that stop elsewhere outside the sample. The fits to 
Eq.~(\ref{eq:TotalPol}) were performed assuming $\sigma_b$ and the ratio $a_s/a_b$ do not change with temperature.
Examples of TF-$\mu$SR spectra and the fits are shown in Fig.~\ref{fig1}. 

\begin{figure}
\centering
\includegraphics[width=8.0cm]{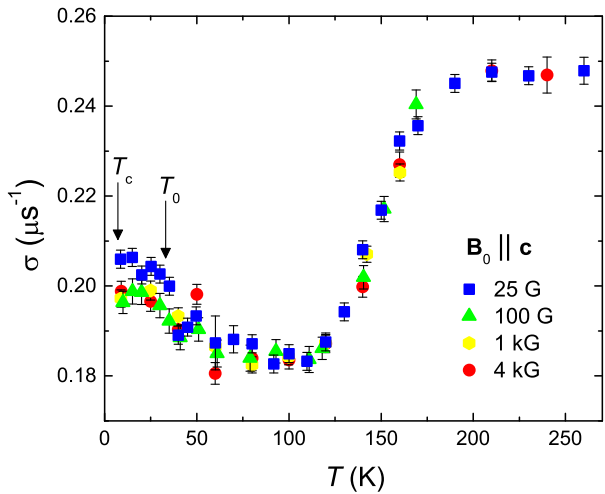}
\caption{(Color online) Temperature dependence of the muon depolarization rate $\sigma$ above $T_c$ for
a magnetic field applied parallel to the $c$ axis. The data for different magnetic field strengths have
been vertically shifted to overlap above $T \! = \! 100$~K.}
\label{fig2}
\end{figure}
 
\begin{figure}
\centering
\includegraphics[width=9.0cm]{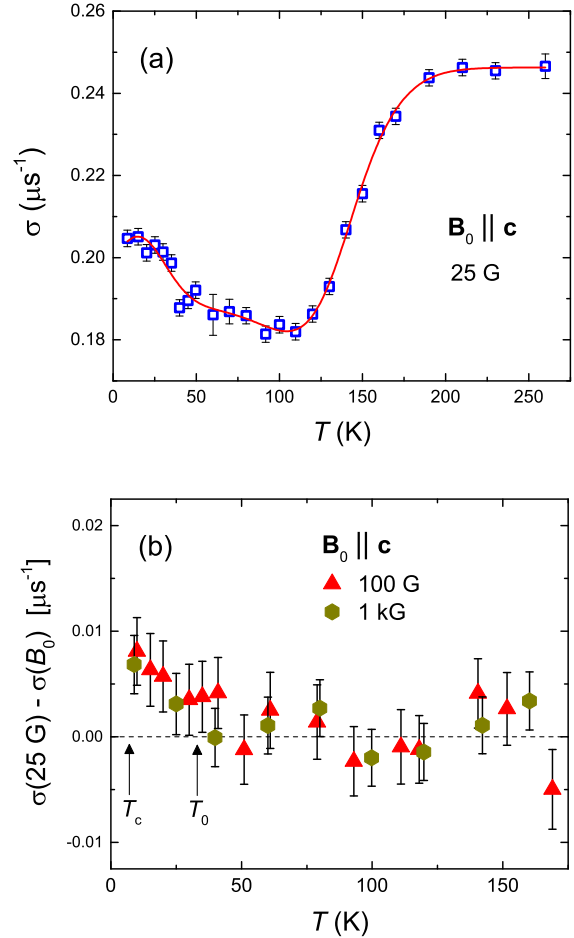}
\caption{(Color online) (a) Temperature dependence of the muon depolarization rate $\sigma$ above $T_c$ for a magnetic field of 25~G 
applied parallel to the $c$ axis (This data is also shown in Fig.~\ref{fig2}). The solid curve through the data points is a guide to the eye.
(b) Difference between $\sigma$ for $B_0 \! = \! 25$~G and 100~G, and for $B_0 \! = \! 25$~G and 1~kG. 
The solid curve from (a) was used to calculate these differences at temperatures where no 25~G data exist.}
\label{fig3}
\end{figure}

\section{Results and Discussion}

Figure~\ref{fig2} shows the temperature dependence of the depolarization rate $\sigma$ above $T_c$ for different magnetic fields 
applied parallel to the $c$ axis ({\bf B}$_0 \! \parallel \!$~{\bf c}). There is some variation of $\sigma$ with the applied
field strength due to some uncertainty in determining the size of the background signal, which has a mean precession frequency 
close to that of the sample signal. Consequently, the data sets for different fields have been vertically shifted in Fig.~\ref{fig2} 
to overlap above $T \! = \! 100$~K, where CDW correlations are not expected. The ratio of the quadrupolar and Zeeman interaction
energies of the $^{93}$Nb nuclei for $B_0 \! = \! 25$~G, 100~G, 1~kG, and 4~kG are $\nu_{\rm Q}/\nu_{\rm Z} \! = \!  99$, 25, 2.5, and 0.6
respectively. Hence the quadrupolar interaction is relevant to a varying degree for all magnetic fields considered in our experiments.    

The strong temperature dependence of $\sigma$ is indicative of thermally-activated hopping of the $\mu^+$. 
With increasing temperature $\sigma$ first decreases, likely due to the $\mu^+$ averaging over 
the distribution of fields it experiences during its lifetime as it moves from site to site (a situation analogous to motional 
narrowing of the line width in NMR). Above $T \! \sim \! 110$~K the increase of $\sigma$ suggests that at these temperatures at least some
of the $\mu^+$ move fast enough to reach sites near defects within their various lifetimes. Consequently, the enhanced $\sigma$ results
from the muon ensemble sensing a broadened distribution of time-averaged magnetic fields. The plateau appearing above $T \! \sim \! 200$~K
suggests that at these higher temperatures the $\mu^+$ hop rate becomes very fast, such that most of the $\mu^+$ reach the vicinity of a defect 
very early in their lifetime --- where they presumably experience a broader distribution of nuclear dipole fields. 
The muon hop rate is not influenced by the applied magnetic field, and hence the behavior of $\sigma$
as a function of temperature is more or less the same for all fields.

\begin{figure}
\centering
\includegraphics[width=8.0cm]{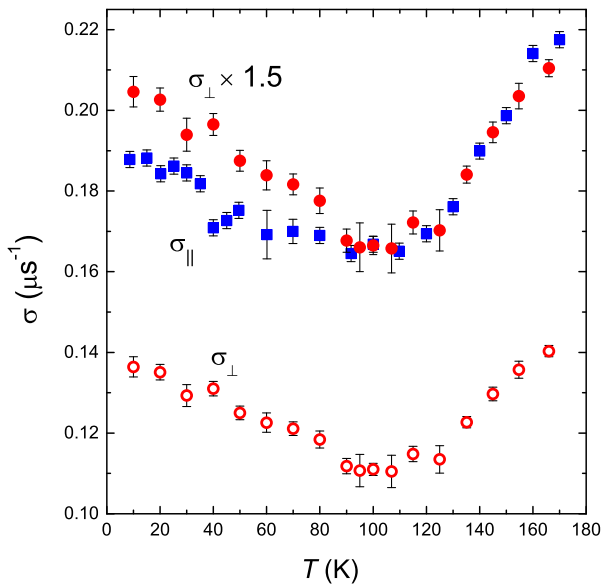}
\caption{(Color online) Temperature dependence of the muon depolarization rate above $T_c$ for a magnetic field of $B_0 \! = \! 25$~G
applied parallel ($\sigma_{\parallel}$) or perpendicular ($\sigma_{\perp}$) to the $c$ axis. The solid red circles are the 
$\sigma_{\perp}$ data multiplied by a scaling factor of 1.5.}
\label{fig4}
\end{figure}
 
In actuality, below $T \! \sim \! T_0$ there is some difference between $\sigma$ for $B_ 0 \! = \! 25$~G and the data for higher
applied fields. While this is noticeable in Fig.~\ref{fig2}, it is more obvious in a plot of the difference between $\sigma$
values. As shown in
Fig.~\ref{fig3}(b), the difference between $\sigma$ for $B_ 0 \! = \! 25$~G and data collected for $B_ 0 \! = \! 100$~G or 1~kG 
increases with decreasing temperature below $T \! \sim \! T_0$. Changes in the form of the ZF-$\mu$SR signal are also observed 
below $T_0$, where there is no {\bf B}$_0$ to affect the $^{93}$Nb quantization axis.\cite{Higemoto:99} 
At $B_0 \! =  \! 25$~G, where the ratio $\langle I_x^{\rm Nb} \rangle/\langle I_z^{\rm Nb} \rangle 
\! \sim \! \nu_{\rm Q}/\nu_{\rm Z}$ is quite large, the quantization axis for the $^{93}$Nb nuclear spin states is also solely
determined by the direction of the maximal EFG at the nuclear site. At all temperatures there are contributions to the net EFG 
at the $^{93}$Nb nuclear site from the crystal lattice and the electric field directed radially away from the $\mu^+$. The onset 
of CDW order below $T_0$ introduces an additional contribution that rotates the direction of the maximal EFG, 
hence changing $\sigma_{\rm Nb}$. Nevertheless, for this orientation of {\bf B}$_0$ the effect is quite weak. 
 
Figure~\ref{fig4} shows the effect of changing the direction of the applied field to be perpendicular to the $c$ axis
({\bf B}$_0 \! \perp \!$~{\bf c}). To maintain the transverse-field geometry, this also requires changing the direction 
of {\bf P}(0) to be parallel (rather than perpendicular) to the $c$ axis. Due to the anisotropic nature of the
dipolar interaction between the $\mu^+$ and the host nuclei, the fitted values of $\sigma$ are expected to show a dependence 
on the change in directions of {\bf B}$_0$ and {\bf P}(0). In particular, the change in direction of {\bf B}$_0$ results in a change in the angles
$\theta_i$ and $\theta_j$ in Eqs.~(\ref{eq:Se}) and (\ref{eq:Nb}), and the change in direction of {\bf P}(0)
corresponds to a change in the direction of the $\mu^+$ spin. For $B_0 \! = \! 25$~G, the quantization axis 
for the $\mu^+$ spin states rotates with {\bf B}$_0$, but the quantization axis for the $^{93}$Nb spin states determined solely
by the net EGF does not. With the direction of the maximal EGF at the $^{93}$Nb nuclear site fixed and continuing to define the 
direction of {\bf B}$_0$ as the $z$ direction, the 90$^{\circ}$ rotation of {\bf B}$_0$ changes the ratio of 
$\langle I_x^{\rm Nb} \rangle/\langle I_z^{\rm Nb} \rangle$ in Eq.~(\ref{eq:Nb}).    

To see the effect on $\sigma$ of changing the angle between {\bf B}$_0$ and the EFG associated with CDW order, we show 
the data for {\bf B}$_0 \! \perp \!$~{\bf c} ({\it i.e.} $\sigma_{\perp}$) in Fig.~\ref{fig4} multiplied by a scaling 
factor that eliminates the anisotropy in $\sigma$ at high temperatures. Below $T \! \sim \! 90$~K there is a clear enhancement of the scaled 
$\sigma_{\perp}$ data above $\sigma_{\parallel}$, indicating a change in the angle between the quantization axes of the
$^{93}$Nb nuclear and $\mu^+$ spin states. This suggests that the $\mu^+$ is sensitive to static CDW order up to temperatures 
nearly 3 times the CDW phase transition temperature $T_0$ --- consistent with the STM study of Ref.~\onlinecite{Arguello:14}, which 
as mentioned earlier shows nanometer-size regions of static CDW order in the vicinity of defects at the surface of 
the sample in the temperature range $T_0 \! < \! T \! < 3 T_0$.
The STM measurements demonstrate an increase of the correlation length of the CDW regions as the temperature is lowered 
toward $T_0$. Assuming the same occurs in the bulk, with decreasing temperature a greater number of the mobile $\mu^+$ 
encounter regions of CDW order during their lifetime.   

\section{Summary}

In the present work we have demonstrated the sensitivity of TF-$\mu$SR to static CDW order in the bulk of 2$H$-NbSe$_2$.
The results support early NMR findings, and recent STM measurements at the surface of 2$H$-NbSe$_2$ showing the 
occurrence of nanodomains of static CDW order above $T_0$ near defects. It is shown that the $\mu^+$ is mobile at temperatures above $T_c$,
and may become trapped by defects during its lifetime. Above $T_0$ this mobility presumably allows a greater number of the 
implanted muons to experience the short-range CDW correlations.    

\begin{acknowledgments}    
We thank R.F. Kiefl for informative discussions and the staff of TRIUMF's Centre for Molecular and Materials Science 
for technical assistance. JES acknowledges support from the Canadian Institute for Advanced Research and 
the Natural Sciences and Engineering Research Council of Canada. 
\end{acknowledgments}

\end{document}